\documentclass[aoas,MSNbibl,nameyear,rotating,dvips]{arximspdf}
\usepackage{graphicx}

\doi{10.1214/12-AOAS540}
\volume{6}
\issue{4}
\pubyear{2012}
\firstpage{1377}
\lastpage{1405}

\makeatletter
\newcommand{\matR}[1]{{\mathbb R}^{#1}}
\newcommand{\PP}{{\mathbb P}}
\newcommand{\EE}{{\mathbb E}}

\newcommand{\tr}{\operatorname{tr}}

\newcommand{\N}{\mathrm{N}}
\newcommand{\diag}{\operatorname{diag}}

\newcommand{\vecepsilon}{\boldsymbol{\varepsilon}}
\newcommand{\vectheta}{\boldsymbol{\theta}}

\newcommand{\vecmu}{\boldsymbol{\mu}}
\newcommand{\vectau}{\boldsymbol{\tau}}
\newcommand{\vecPsi}{\boldsymbol{\Psi}}
\newcommand{\vect}{\mathbf{t}}
\newcommand{\vecW}{\mathbf{W}}
\newcommand{\vecy}{\mathbf{y}}
\newcommand{\vecI}{\mathbf{I}}
\newcommand{\vecS}{\mathbf{S}}
\newcommand{\vecSigma}{\boldsymbol{\Sigma}}
\newcommand{\vecK}{\mathbf{K}}
\newcommand{\vecD}{\mathbf{D}}
\newcommand{\vecG}{\mathbf{G}}

\def\vec0{{\mathbf0}}

\makeatother

\begin{document}
\begin{frontmatter}

\title{Finding a consensus on credible features among several
paleoclimate reconstructions}
\runtitle{Consensus of paleoclimate reconstructions}

\begin{aug}
\author[A]{\fnms{Panu} \snm{Er\"{a}st\"{o}}\ead[label=e1]{panu.erasto@helsinki.fi}},
\author[B]{\fnms{Lasse} \snm{Holmstr\"{o}m}\corref{}\ead[label=e2]{lasse.holmstrom@oulu.fi}},
\author[C]{\fnms{Atte} \snm{Korhola}\ead[label=e3]{atte.korhola@helsinki.fi}}
\and\\
\author[C]{\fnms{Jan} \snm{Weckstr\"{o}m}\ead[label=e4]{jan.weckstrom@helsinki.fi}}
\runauthor{Er\"{a}st\"{o}, Holmstr\"{o}m, Korhola and Weckstr\"{o}m}
\affiliation{National Institute for Health and Welfare and University
of Oulu,
University of~Oulu,
University of Helsinki and
University of Helsinki}
\address[A]{P. Er\"{a}st\"{o}\\
National Institute for Health\\ \quad  and Welfare\\
P.O. Box 30, FIN-00271 Helsinki\\
Finland\\ \printead{e1}}
\address[B]{L. Holmstr\"{o}m\\
Department of Mathematical Sciences\\University of Oulu\\
P.O. Box 3000, FIN-90014\\
Finland\\ \printead{e2}}
\address[C]{A. Korhola\\
J. Weckstr\"{o}m\\
Environmental Change Research Unit (ECRU)\\
Department of Environmental Sciences\\
University of Helsinki\\
P.O. Box 65, FIN-00014\\
Finland\\ \printead{e3}\\
\hphantom{\textsc{E-mail}:\ }\printead*{e4}} 
\end{aug}

\received{\smonth{5} \syear{2011}}
\revised{\smonth{11} \syear{2011}}

%
\begin{abstract}
We propose a method to merge several paleoclimate time series into one
that exhibits a consensus on the features of the individual times
series. The paleoclimate time series can be noisy, nonuniformly sampled
and the dates at which the paleoclimate is reconstructed can have
errors. Bayesian inference is used to model the various sources of
uncertainty and smoothing of the posterior distribution of the
consensus is used to capture its credible features in different time
scales. The technique is demonstrated by analyzing a collection of six
Holocene temperature reconstructions from Finnish Lapland based on
various biological proxies. Although the paper focuses on paleoclimate
time series, the proposed method can be applied in other contexts where
one seeks to infer features that are jointly supported by an ensemble
of irregularly sampled noisy time series.
\end{abstract}

%

\end{frontmatter}

\section{Introduction}\label{sec1}

Paleoclimatological proxy data, such as pollen, tree rings or ice
cores, considered to be sensitive to past surface temperature
variations can provide a continuous and long record of climatic changes
where long-term instrumental data are lacking
[\citet{Janetal2007}]. Paleoclimatological data are essential to place
limited instrumental records in perspective and to assess the
importance of forcing factors. However, it is important to realize that
proxy records are indirect measures of climate change that often
reflect changes in multiple aspects of climate [e.g., \citet{LeG2006}; \citet{Tinetal10}]. Each proxy inevitably has its advantages and
limitations, and different proxies may yield information on different
aspects of climate. For example, they may be sensitive to different
seasonal signals, have different response times, and respond directly
or indirectly to climate.
It is therefore not surprising that, for example, temperature
reconstructions based on different proxies can produce somewhat
different results, despite the fact that they reflect a common
underlying truth. One would therefore like to have a method that could
capture, in a principled manner, those aspects of different
reconstructions that find strongest support among most of them, that
is, establish a ``consensus'' on the underlying features of the reconstructions.

\begin{figure}

\includegraphics{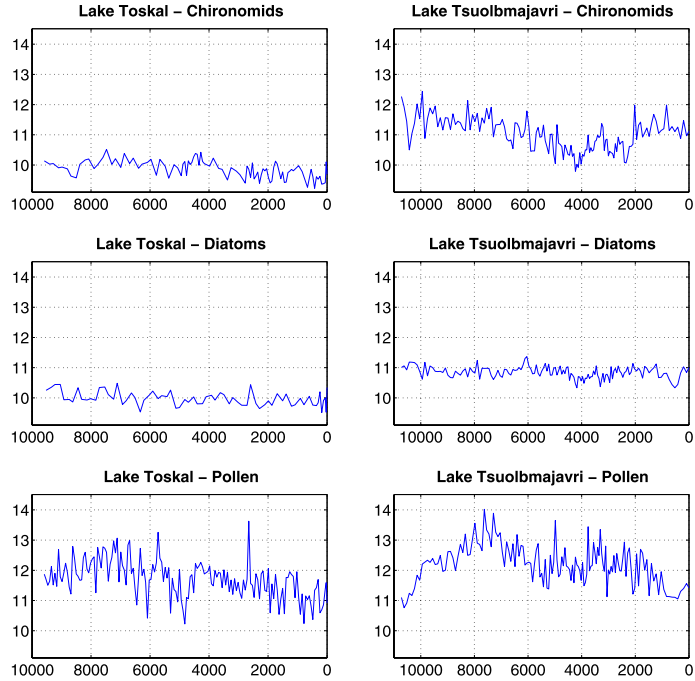}

\caption{The six Holocene mean air July temperature reconstructions
for Northern Fennoscandia used in the consensus analysis. The vertical
axes show temperature in centigrade ($^{\circ}$\textup{C}) and the horizontal axes are
calibrated years before present.%
}
\label{all6reconstructions}
\end{figure}

\begin{figure}

\includegraphics{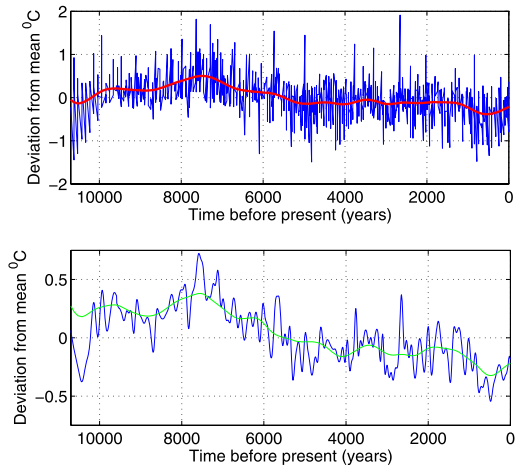}

\caption{Simple methods to establish a consensus between temperature
reconstructions. Upper panel: all six reconstructions of Figure~\protect\ref
{all6reconstructions} centered and stacked together (blue) and a local
linear regression smooth (red). Lower panel: averages of cubic spline
interpolants (blue) and local linear regression smooths (green) of the
centered reconstructions. Local linear regression smooths employ a
Gaussian kernel and bandwidths computed using a method from \protect\citet{ruppert+sheather+wand1995}.}
\label{simplemethods}
\end{figure}

To demonstrate the method suggested in this paper, we will find a
consensus among the six
Holocene, that is,  post Ice Age mean air July
temperature reconstructions shown in Figure~\ref{all6reconstructions}.
The reconstructions are based on three biological proxies analyzed from
two lakes in Finnish Lapland and,
as one can see, they differ from one another considerably, both in the
overall temperature levels and in the details. The data behind the
reconstructions and the consensus features the proposed method finds
will be discussed in detail in Section~\ref{application}, but let us
first consider here some
%
ad hoc methods that are often used to combine information across these
types of paleoclimate time series. Such straightforward analyses are
demonstrated in Figure~\ref{simplemethods}. In the upper panel the
reconstructions have been centered and then stacked into a single plot.
A smooth has also been computed and it can be interpreted to represent
the consensus temperature anomaly, that is, deviation from mean. In the
lower panel the centered reconstructions have been averaged after first
interpolating them with cubic splines or, alternatively, by smoothing
them with local linear regression.
While simple plots like these may reveal some features of the consensus
anomaly, they clearly leave many questions unanswered.
Individual time series are noisy, as both the reconstructed
temperatures and the dates they are thought to correspond to contain errors.
Such
simple methods also tell us nothing about the uncertainty in the
suggested consensus features that the presence of noise inevitably introduces.
Further, the underlying signal may exhibit interesting features in many
different time scales and a single smooth or mean probably cannot
capture all of them well.

In climate science, a popular approach to reconstruct large-scale past
climate variation is to combine a number of individual proxy records
using the so-called Composite Plus Scaling (CPS) method [e.g., Jones et
al.
(\citeyear{Jonetal09}) and the references therein)]. In this method, a collection of
proxy records is standardized and averaged after which the average is
recalibrated against an available instrumental record of a particular
environmental variable, such as temperature. In the calibration
process, various regression techniques can be used to match an average
of annually resolved proxy records with modern instrumental data. The
method proposed in this paper works differently in that the individual
reconstructions are not explicitly standardized or averaged and their
consensus is found using an estimation process that does not directly
rely on a modern instrumental record. Note that, contrary to the
situation with annually resolved proxies such as tree rings, in the
case of biological proxy records considered here only a few of the
reconstructed temperatures would fall in a period for which
instrumental measurements might be available, making regression based
calibration unfeasible.

Our proposal to consensus analysis is a Bayesian approach that consists
of two steps. First, given a set of reconstructions, we find their
consensus by viewing the reconstructions as data in a hierarchical
model that takes into account the uncertainties involved. In the second
step we use scale space smoothing to reveal the salient features of the
consensus in different time scales. The proposed approach was first
outlined in \citet{nonrefKoretal06b} and
\citet{nonrefHoletal08} and it can be viewed as an extension to
multiple time series of the BSiZer methodology that has already found
use in quantitative paleoecological analyses
[\citeauthor{refEraHol03} (\citeyear{refEraHol03,refEraHol05b,refEraHol05}); \citet{refHol09b}; \citet{refWecetal05}].

It can be argued that a better way to model the propagation of errors
into the consensus would be to work directly with Bayesian temperature
reconstructions instead of using a Bayesian model to combine
non-Bayesian reconstructions, as is done here. However, while Bayesian
models may be becoming more commonplace, the vast majority of existing
reconstructions are in fact non-Bayesian, based on various regression
techniques, both parametric and nonparametric. See,
for example, \citet{Bir95} and \citet{Biretal10} for extensive
reviews of the kind of methods typically used in connection with
diatoms, pollen, chironomids and other biological proxies. The method
proposed here is therefore immediately widely applicable as a
significant improvement over the simplistic ad hoc summaries commonly
used to represent a consensus of such reconstructions.

To our knowledge, the first papers to describe a detailed Bayesian
modeling approach to biological proxy based paleoclimate reconstruction
are \citet{Vasetal00}, \citet{Toietal01} and \citet{Koretal02}, who
all used chironomid taxon abundances in lake sediments as temperature
proxy. Their approach was further analyzed by Er\"{a}st\"{o} and
Holmstr\"{o}m (\citeyear{refEraHol05b}) and more recently by \citet{refSaletal11}.
Bayesian reconstruction based on pollen abundances was described in
\citet{Hasetal06}. All these papers model explicitly the response of a
biological proxy to temperature changes and reconstruct the temperature
from taxon fossil abundance data in a single proxy record. More
recently, a Bayesian hierarchical model was used by \citet{BryBer11}
to reconstruct climate for the past 400 years from several bore hole
temperature profiles.

The approach suggested in \citet{Lietal10} is perhaps closer to the
one proposed here in that a number of local reconstructions are
combined to create a single temperature reconstruction, in their case
for the whole northern hemisphere and the last 1000 years. As in the
present paper, a biological proxy (pollen) enters the reconstruction
process only as a temperature time series and not as raw taxon
abundances, which would constitute the original data. In addition to
pollen, tree rings and bore hole temperatures are also used in their
model and external forcings are accounted for as well. However, no real
proxy data are used and instead the proxy records are simulated on the
basis of numerical climate model outputs. The reconstructions we aim to
combine were obtained using taxon abundance data from actual sediment
cores. Note that the same climate model simulation that was used in
\citet{Lietal10} is employed also in the present paper but only to
elicit a prior density for the consensus reconstruction. Other
differences include the somewhat more general error models considered
here, explicit modeling of dating uncertainty and the scale space
approach to inference.

In Section~\ref{method} we describe our method, assuming first fixed
dates for the reconstructed temperatures (Section~\ref{fixeddates})
and then allowing dating errors in the analysis (Section~\ref
{randomdates}). The idea of using multi-scale smoothing to capture
temperature variation in different time scales is explained in
Section~\ref{featureanalysis}. The analysis of the consensus features
in the six Holocene temperature reconstructions is presented
in Section~\ref{application} and Section~\ref{discussion} offers a
discussion of the main points of the paper.
The Matlab functions used in the main computations are provided in
\citet{Eraetal11suppb}.

\section{The method}\label{sec2}
\label{method}

\subsection{Fixed dates}\label{sec2.1}
\label{fixeddates}

The method that we will describe can be used to analyze reconstructions
of any continuous variable, but as our main interest is in the Holocene
temperature, we frame the following description in terms of temperature
reconstructions. Thus,
consider $m$ reconstructions $\vecy_1,\ldots,\vecy_m$ of past temperatures,
where $\vecy_k = [y_{k1},\ldots,y_{kj_k}]^T$ are the estimated past
temperatures from the $k$th proxy series and let
$\vect_k = [t_{k1},\ldots,t_{kj_k}]$ be the associated radiocarbon
dating based chronology. Here $t_{k1} < \cdots< t_{kj_k}$ so that
$y_{k1}$ and $y_{kj_k}$ are the reconstructions for the oldest and the
youngest dates, respectively.
We assume that the reconstructions are from a relatively limited
geographical area so that they can be thought to reflect common
underlying temperature variation and it is this common variation that
we seek to capture.

In the example we will consider the reconstructions are based on fossil
records in sediment cores obtained from subarctic lakes.
Even when the cores come from a limited area, due to, for example,
different lake altitudes, the overall temperature levels and therefore
the mean temperatures in the reconstructions can vary considerably. We
therefore consider only temperature anomalies, centering each
reconstruction $\vecy_k$ by subtracting its mean $(1/j_k)\sum_l^{j_k}
y_{kl}$ from all components $y_{kl}$. These centered time series
represent reconstructions of past temperature anomalies (variation
about the mean) and we attempt to capture the statistically significant
(or ``credible'') features in what can be interpreted as the consensus
of these anomalies in the general area where the core lakes are
located. The features in the consensus that we are interested in are
locations of maxima, minima and trends, all of which are not affected
by centering. To avoid the introduction of new notation, we denote the
centered reconstructions still by $\vecy_k$.

The consensus anomaly is modeled as a curve $\mu(t)$, where $t \in
[a,b]$ is a time interval that includes all chronologies from all proxy
records. We actually assume that $\mu$ can be described by a natural
cubic spline with knots at the points $t_{kj_l}$. Such a spline is
uniquely determined by its values at the knots because they determine
the interpolating spline uniquely [\citet{GreSil94}]. The fact that
this spline space is finite dimensional greatly simplifies our analysis.

Let
%
\begin{equation}
\label{obsdates}
\vect= \{t_1,\ldots,t_n\} = \bigcup_k^m \{t_{k1},\ldots,t_{kj_k}\}
\end{equation}
be the set of distinct dates, in increasing order,
in all chronologies $\vect_k$. Since all $t_{kl}$'s need not be
different, we have that
$n \leq j_1+\cdots+ j_m$. The anomaly curve is modeled as a natural
cubic spline with values $\mu_i = \mu(t_i)$ at the knots $t_i$. Thus,
instead of $\mu$, we can from now on work with the finite dimensional
vector $\vecmu= [\mu_1,\ldots,\mu_n]^T$ of past anomalies at times $t_i$.

Now, let $\vecmu_k$ be the part of $\vecmu$ that corresponds to the
chronology $\vect_k$ of
the $k$th reconstruction $\vecy_k$. We assume that
%
\begin{equation}
\label{model1}
\vecy_k = \vecmu_k + \vecepsilon_k,
\end{equation}
where $\vecepsilon_k$ has the multivariate normal distribution $\N
(\vec0,\vecSigma_k)$ with an unknown covariance matrix $\vecSigma
_k$. Our model therefore allows time-varying, correlated reconstruction
errors that can also have different magnitudes for different proxies
and cores. Such a model is supported by the exploratory analysis
reported in
\citet{Eraetal11supp}. We further assume that the anomalies are
conditionally independent given the parameters $\vecmu$ and $\{
\vecSigma_k\} = \{\vecSigma_1,\ldots,\vecSigma_m\}$ so that the
likelihood of the data $\vecy= [\vecy_1^T,\ldots,\vecy_m^T]^T$,
given these parameters, is
%
\begin{equation}
\label{likelihood1}
p(\vecy| \vecmu,\{\vecSigma_k\}) \propto\prod_{k=1}^m |\vecSigma_k|^{-1/2}
\exp \biggl[-\frac{1}{2}(\vecy_k - \vecmu_k)^T \vecSigma
_k^{-1}(\vecy_k - \vecmu_k) \biggr].
\end{equation}

As a prior distribution for $\vecSigma_k$ we use an Inverse Wishart
distribution,
%
\begin{equation}
\label{wishartprior}
p(\vecSigma_k | \vecW_k,\nu_k) \propto
|\vecSigma_k| ^{-(\nu_k+j_k+1)/2} \exp \bigl[-\tfrac{1}{2}\tr
(\vecW_k\vecSigma_k^{-1} ) \bigr],
\end{equation}
a standard choice in connection with a multivariate normal likelihood.
As there seldom is any prior knowledge of a particular error
correlation structure, we typically use a diagonal prior scale matrix
$\vecW_k$ and select the degrees of freedom $\nu_k$ so that the prior
(\ref{wishartprior}) is rather vague, allowing nondiagonal posterior
covariances. The relative magnitudes of the diagonal elements of $\vecW
_k$ could also be used to model the increased level of difficulty of
temperature reconstruction for the older sediment layers [\citet{refEraHol05}]. The $\vecSigma_k$'s are assumed to be independent  a priori
 so that
%
\begin{equation}
\label{indepsigmas}
p(\{\vecSigma_k\}) = p(\{\vecSigma_k\} | \{\vecW_k,\nu_k\}) =
\prod_{k=1}^m p(\vecSigma_k | \vecW_k,\nu_k).
\end{equation}

We have also experimented with a more complex model that allows
reconstruction error correlations between different proxy records. Let
again $\vecy= [\vecy_1^T,\ldots,\vecy_m^T]^T$
be the vector of length $j_1+\cdots +j_m$ that contains all
reconstructions. The more complex model considered assumes that
%
\begin{equation}
\label{extlikelihood}
p(\vecy| \vecmu,\vecSigma) \propto|\vecSigma|^{-1/2}
\exp \bigl[-\tfrac{1}{2}(\vecy-\vecG\vecmu)^T \vecSigma^{-1}(\vecy
-\vecG\vecmu) \bigr],
\end{equation}
where $\vecG\vecmu$ is a modification of the consensus $\vecmu$
where some components $\mu_i$ appear several times to account for the
fact they correspond to dates in the joint chronology that appear in
more than one reconstruction. The covariance matrix $\vecSigma$ again
has an inverse-Wishart prior
%
\begin{equation}
\label{wishartprior2}
p(\vecSigma| \vecW,\nu) \propto
|\vecSigma| ^{-(\nu+j+1)/2} \exp \bigl[-\tfrac{1}{2}\tr (\vecW
\vecSigma^{-1} ) \bigr],
\end{equation}
where now $j=j_1+\cdots +j_m$ and $\vecW$ is the diagonal matrix whose diagonal
elements are those of the matrices $\vecW_1,\ldots,\vecW_m$. The
results reported in the paper all pertain to the model (\ref
{likelihood1}) and the more complex model (\ref{extlikelihood}) is
discussed in
\citet{Eraetal11supp}.

For the consensus anomaly $\vecmu$ we use a smoothing prior that
penalizes for roughness as measured by the variability of its components,
%
\begin{equation}
\label{muprior}
p(\vecmu| \lambda_0,\vect) \propto\lambda_0^{(n-2)/2}\exp
\biggl(-\frac{\lambda_0}{2}
\vecmu^T\vecK\vecmu \biggr).
\end{equation}
In this formula, $\vecK$ is a symmetric positive semidefinite matrix
such that
%
\begin{equation}
\label{roughnesspenalty}
\vecmu^T\vecK\vecmu= \int_a^b[\mu''(t)]^2\,dt
\end{equation}
and $\lambda_0 > 0$. Thus, the roughness in the prior (\ref{muprior})
is measured by the second derivative of the natural cubic spline that
interpolates the values $\vecmu$ at the knots $t_i$ and the level of
roughness penalty is controlled by $\lambda_0$
[\citet{GreSil94}]. The power
$(n-2)/2$ in the scaling factor reflects the rank of the matrix $\vecK
$ which is $n-2$.
Note that the smoothing prior (\ref{muprior}) imposes dependence
between the temperature anomalies $\vecmu_k$ derived from these
proxies. This is natural because the reconstructions are assumed to reflect
common underlying temperature variation.

The parameter $\lambda_0$ describes our prior beliefs about the
smoothnesss of~$\mu$. We consider it unknown with prior uncertainty
described by a Gamma distribution. In principle, point estimation such
as cross-validation can be used to choose suitable values for the prior
distribution parameters [\citet{refEraHol03}], but we prefer here a
choice that produces a posterior mean of $\mu$ of reasonable roughness.
The important thing is to avoid choosing $\lambda_0$ too large because
then the finest details of $\mu$ might be lost [\citeauthor{refEraHol03} (\citeyear{refEraHol03,refEraHol05})].


The joint posterior distribution of all the unknown parameters in the
model is now obtained from the Bayes' formula,
%
\begin{equation}
\label{fullposterior}
p(\vecmu,\{\vecSigma_k\},\lambda_0 | \vecy, \vect) \propto
p(\lambda_0)p(\{\vecSigma_k\})p(\vecmu| \lambda_0,\vect)p(\vecy|
\vecmu,\{\vecSigma_k\}),
\end{equation}
where all the distributions on the right-hand side were defined above.
Gibbs sampling can be used to generate a sample from this posterior
distribution.
An estimate of the consensus anomaly that is consistent with the data
and our prior beliefs, together with its uncertainty, is described by
the marginal posterior distribution
$p(\vecmu| \vecy)$, which then can be approximated by
the $\vecmu$-component of this sample.
The model (\ref{extlikelihood}) is handled similarly.


\subsection{Random dates}\label{sec2.2}
\label{randomdates}

In the previous section we assumed that the reconstructed temperature
anomalies $y_{kl}$ could be associated precisely with the dates
$t_{kl}$. In reality, however, the core chronologies are derived from
radiocarbon dating based estimates, a process that is not error-free.
Taking into account this source of uncertainty can be important when
one tries to make inferences about the common features in several
temperature time series with different associated chronologies.

Let $\vect_k = [t_{k1},\ldots,t_{kj_k}]$ again be the radiocarbon
dating based chronology for the $k$th reconstruction. Allowing for the
fact that the dates $t_{kl}$ have errors, we assume that they and the
dates $\tau_{kl}$ in the true, unobserved chronology, satisfy $t_{kl}
= \tau_{kl} + \delta_{kl}$, where $\delta_{kl}$ represents an error.
Denote the true chronology for the $k$th reconstruction by
$\vectau_k = [\tau_{k1},\ldots,\tau_{kj_k}]$. We assume that both
sequences $\vect_k$ and $\vectau_k$ are strictly increasing. Note
that, for $k \neq k'$, $\vectau_k$ and $\vectau_{k'}$ may well
contain some dates that are known to be the same. This is the case, for
example, when $k$ and $k'$ correspond to two different proxies analyzed
from the same core and using the same sediment samples for both. Let
%
\begin{equation}
\label{truedates}
\vectau= \{\tau_1,\ldots,\tau_n\} = \bigcup_k^m \{\tau
_{k1},\ldots,\tau_{kj_k}\}
\end{equation}
be the set of distinct dates in all chronologies $\vectau_k$,
$k=1,\ldots,m$
[cf. (\ref{obsdates})]. As with the dates $t_{kl}$ in the previous
section, since all $\tau_{kl}$'s need not be different, we have in
general that $n \leq j_1+\cdots+ j_m$. The observed dates $t_{kl}$ for
equal $\tau_{kl}$'s are assumed to be also equal and we denote by
$\vect= \{t_1,\ldots,t_n\}$ the set of
$t_{kl}$'s corresponding to~$\vectau$. Our model for these distinct
dates now is
%
\begin{equation}
\label{datingerrormodel}
t_i = \tau_i + \delta_i,
\end{equation}
$i = 1,\ldots,n$, and we assume that, given the parameters $\tau_i$,
the $\delta_i$'s are independent zero mean normal variables with known
variances $\psi_i^2>0$. The variances that we will use are based on
the standard errors associated with the chronologies (cf. Section~\ref
{chronerrors}). The likelihood of the observed dates $\vect$ from
(\ref{datingerrormodel}) is
%
\begin{equation}
\label{datelikelihood}
p(\vect| \vectau) = p(\vect| \vectau,\{\psi_i^2\}) \propto
\prod_{i=1}^n \psi_i^{-1}\exp \biggl[-\frac{1}{\psi_{i}^2}(t_{i} -
\tau_{i})^2 \biggr],
\end{equation}
where $\{\psi^2_i\} = \{\psi_1^2,\ldots,\psi_n^2\}$. We set a prior
distribution on the
$\tau_i$'s that enforces the correct temporal order of the chronology
within each reconstruction,
%
\begin{equation}
\label{tauprior}
p(\vectau) \propto\prod_{k=1}^m 1(\tau_{k1} < \tau_{k2}< \cdots
<\tau_{kj_k}).
\end{equation}

Let now $\tau_{(1)} < \cdots<\tau_{(n)}$ be a permutation of $\vectau
$ into an ascending order. The consensus anomaly is then modeled as
natural cubic spline $\mu(\tau)$ with knots at the points $\tau
_{(i)}$, uniquely determined by the vector
$\vecmu= [\mu_1,\ldots,\mu_n]^T$, $\mu_i = \mu(\tau_{(i)})$. The
subsequent model details are exactly the same as in the previous
section with the exception that in the prior (\ref{muprior}) of
$\vecmu$, the matrix $\vecK$ now depends on $\vectau$.
The joint posterior (\ref{fullposterior}) becomes
%
\begin{eqnarray}
\label{fullposterior2}
  p(\vecmu,\{\vecSigma_k\},\lambda_0,\vectau| \vecy,\vect) &\propto&
p(\lambda_0)p(\{\vecSigma_k\})p(\vectau)p(\vecmu| \lambda
_0,\vectau)\nonumber
\\[-8pt]
\\[-8pt]
&&{}\times p(\vecy| \vecmu,\{\vecSigma_k\})p(\vect| \vectau).
\nonumber
\end{eqnarray}
A hybrid algorithm that uses Gibbs and Metropolis--Hastings Monte Carlo
sampling can be used to generate a sample from this posterior
distribution [e.g., \citet{RobCas05}].
The proposal density for $\tau_i$ is $\N(0,10^{-2}\psi_i^2)$.
Again, the model (\ref{extlikelihood}) can be handled similarly.
For easy reference, Table~\ref{Glossary} summarizes the quantities
defined in this and the previous section.

%
\begin{sidewaystable}
\tabcolsep=0pt
\tablewidth=\textwidth
\caption{Glossary of symbols used, their associated likelihoods or
priors and the full conditional posteriors of\break the estimated parameters.
The multivariate normal distribution
$\N(\vecmu_0,\vecSigma_0)$ in the conditional posterior of\break $\vecmu$
is obtained as the product of (\protect\ref{likelihood1}) and (\protect\ref{muprior})
and it is discussed in   Appendix~\protect\ref{weightanalysis}.
In the conditional posterior of\break $\vectau$ we denote
$\vecPsi= \diag(\psi_1^2,\ldots,\psi_n^2)$ [cf. (\protect\ref
{datelikelihood})] and
the proposal density for $\tau_i$ is $\N(0,10^{-2}\psi_i^2)$}
\label{Glossary}
\begin{tabular*}{\textwidth}{@{\extracolsep{\fill}}llcl@{}}
\hline
&&\textbf{Likelihood}\\
\textbf{Symbol} & \textbf{Meaning} &  \textbf{or prior} & \textbf{Full conditional posterior} \\
\hline
$\vecy_k$ & reconstructed anomaly for proxy record $k$ & (\ref
{likelihood1}) & \\
$\vecy$ &$[\vecy_1^T,\ldots,\vecy_m^T]^T$ &(\ref{extlikelihood}) &
\\
$\vecmu$ & consensus anomaly & (\ref{muprior}) &
$\vecmu| \{\vecSigma_k\},\lambda_0, \vectau, \vecy, \vect  \sim
  \N(\vecmu_0,\vecSigma_0)$ \\
$\vecmu$ & consensus anomaly (extended model) & (\ref{muprior}) &
$\vecmu| \{\vecSigma\},\lambda_0, \vectau, \vecy, \vect  \sim
\N((\vecG+\lambda_0\vecSigma^{-1}(\vecG^T)^{-1}\vecK)\vecy,
(\vecG^T\vecSigma^{-1}\vecG+\lambda_0\vecK)^{-1})$ \\
$\lambda_0$ & prior smoothing parameter of $\vecmu$ &
$\operatorname{Gamma}(\eta,\beta)$ &
$\lambda_0 | \vecmu,\{\vecSigma_k\}, \vectau, \vecy, \vect  \sim
\operatorname{Gamma} ((n-2)/2 + \eta,
\vecmu^T\vecK\vecmu/2 + \beta )$ \\
$\vecmu_k$ & part of $\vecmu$ corresponding to proxy record $k$ & & \\
$\vecepsilon_k$ & $\vecy_k - \vecmu_k$ & & \\
$\vecSigma_k$ & covariance of $\vecepsilon_k$ & (\ref{wishartprior}) &
$\vecSigma_k   |   \vecmu,\lambda_0,\vectau,\vecy, \vect \sim
\operatorname{Inv\mbox{-}Wishart}_{\nu_k+1}
 ( [(\vecy_k-\vecmu_k) (\vecy_k-\vecmu_k)^T+\vecW
_k]^{-1} )$
\\
$\vecSigma$ & covariance of $[\vecepsilon_1^T,\ldots,\vecepsilon
_m^T]^T$ & (\ref{wishartprior2}) &
$\vecSigma  |   \vecmu,\lambda_0,\vectau,\vecy, \vect \sim
\operatorname{Inv\mbox{-}Wishart}_{\nu+1}
 ( [(\vecy-\vecG\vecmu)(\vecy-\vecG\vecmu)^T+\vecW
]^{-1} )$
\\
$\vect_k$ & chronology for proxy record $k$ & & \\
$\vect$ & set of distinct dates in the chronologies $\vect_k$ & (\ref
{datelikelihood}) & \\
$\vectau_k$ & true chronology for proxy record $k$ & & \\
$\vectau$ & set of distinct dates in the true chronologies $\vectau
_k$ & (\ref{tauprior}) &
$\vectau| \vecmu, \{\vecSigma_k\}, \lambda_0, \vecy, \vect\propto
\exp (-\frac{1}{2}  ( (\vectau-\vect)^T\vecPsi
^{-1}(\vectau-\vect)+
\lambda_0\vecmu^T\vecK\vecmu  )  )p(\vectau)$\\
\hline
\end{tabular*}
\end{sidewaystable}

\subsection{Scale space feature analysis}\label{sec2.3}
\label{featureanalysis}

The two previous sections showed how to estimate the consensus of
several temperature reconstructions. This section explains how to find
its credible features in different time scales. The key idea is that of
a scale space. This concept has its roots in computer vision, but it
has recently inspired a host of new statistical data analysis tools
based on multi-scale smoothing. For an overview of these methods we
refer to \citet{refHol09}.\vadjust{\goodbreak}

In the context of this article, the scale space approach amounts to
using smoothing to make inferences about the credible, or
``statistically significant,'' features of the consensus anomaly $\mu$
underlying the data. Thus, suppose that $S_{\lambda}$ is a smoothing
operator associated with a smoothing level
$\lambda> 0$ and let $\mu_{\lambda} = S_{\lambda}\mu$ be the
corresponding smooth of
$\mu$. In the classical scale space literature [e.g., \citet{Lin94}],
the smoother $S_{\lambda}$ would typically be a Gaussian convolution
(moving average with Gaussian weights) with convolution kernel width
(the averaging window) determined by $\lambda$. However, in the
statistical literature other smoothers are often used.

The idea is to make inferences about the features of
$\mu_{\lambda}$ for a range of smoothing levels $\lambda$. Each $\mu
_{\lambda}$ is interpreted to reveal features of $\mu$ at a certain
time scale, little smoothing (small
$\lambda$) showing the short time scale variation and heavy smoothing
(large $\lambda$) revealing the coarsest features, such as the overall
trend. We are, in particular, interested in the maxima and minima of
$\mu_{\lambda}$ and therefore base our inferences on the derivative
$\mu_{\lambda}'$ because its sign tells where the local trend is
positive or negative.
For Bayesian reasoning we need the posterior $p(\mu_{\lambda}' |
\vecy,\vect)$. However, as the spline $\mu$ is uniquely represented
by the vector
$\vecmu$ of its values at the knots, we may instead consider a
smoothing matrix $\vecS_{\lambda}$, the smooth $\vecmu_{\lambda} =
\vecS_{\lambda}\vecmu$, and then use another matrix $\vecD$ [e.g.,
\citet{GreSil94}] to evaluate the derivative $\mu_{\lambda}'$ at some
fixed dense set of time points $ s_1<\cdots<s_r$,
%
\begin{equation}
\label{derivatives}
\vecD\vecmu_{\lambda} = [\mu_{\lambda}'(s_1),\ldots,\mu_{\lambda
}'(s_r)]^T.
\end{equation}

The smoothing matrix used in our scale space feature analysis is
defined as
$\vecS_{\lambda} = (\vecI+ \lambda\vecK)^{-1}$ and it actually
smooths a discrete set of points $\vecmu$ by fitting a smoothing
spline [\citet{GreSil94}]. Instead of $p(\mu_{\lambda}' | \vecy
,\vect)$, one can now analyze the posterior distribution
$p(\vecD\vecS_{\lambda}\vecmu| \vecy,\vect)$. For fixed dates, a
large sample can first be generated from $p(\vecmu| \vecy,\vect)$
and then transformed by multiplying the sample vectors by the matrix
$\vecD\vecS_{\lambda}$. Inference about the features of $\mu$ at
the time scale $\lambda$ is then based on this sample. With random
dates, the scale space analysis needs samples from both $\vecmu$ and
$\vectau$, as the smoothing matrix $\vecS_{\lambda}$ depends on
$\vectau$ through $\vecK$.

Note here the difference between the parameter $\lambda_0$ used in
constructing the consensus and the parameter $\lambda$ in scale space
feature analysis: $\lambda_0$ describes our prior beliefs about the
underlying consensus $\mu$, whereas different values of $\lambda$ are
used to explore the features of $\mu$ in different time scales.
The choice of prior distribution for $\lambda_0$ is discussed in
Section~\ref{roughness}.
We also emphasize that all inferences on the features of $\mu$ are
made in a simultaneous fashion, over all time points $s_j$ in (\ref
{derivatives}). Therefore, instead of just examining the statistical
significance of individual slopes $\mu(s_j)$, the credibility of whole
patterns of trends are established. For more details on the inference
procedures used we refer to \citet{refEraHol03}.

\section{Holocene temperature variation in Finnish Lapland}\label{sec3}
\label{application}
\subsection{The data used}\label{sec3.1}
\label{data}

We demonstrate the proposed method by finding the consensus among
six temperature reconstructions based on high resolution lake
sedimentary data (50--70 year intervals) of three biological proxies
from two sites (Figure~\ref{all6reconstructions}).
The
two lakes, Toskal and Tsuolbmajavri, selected for analysis are located
at a climatically sensitive tree-line region of Finnish Lapland. They
both contain fossil records of three fundamental climate proxies,
pollen, chironomids (nonbiting midges) and diatoms (unicellular
micro-algae) from the same sediment cores.
The sediments of such remote lakes at high altitudes and latitudes are
perhaps one of the few systems where a continuous, high resolution
record of terrestrial environmental variability, uninfluenced by human
impact throughout the post-glacial, can be found.

Past temperatures were reconstructed using regional training sets of
lakes for pollen, chironomids and diatoms (304, 62 and 64 lakes,
resp.) and a regression based reconstruction technique referred
to as weighted averaging partial least squares (WA-PLS) [\citet{Braaetal93}]. The model relates the modern mean July temperatures at
the training lakes to the abundances of various proxy taxa preserved in
the top (0--1 cm) surface sediments that represent the last few years
of sediment accumulation. The past air temperatures are reconstructed
by substituting in the regression model the taxon abundances found in
the sediment cores from the two lakes selected for analysis. This
approach is based on the assumption that each taxon has a certain
optimal temperature at which it fares particularly well and that,
therefore, the relative abundances of taxon fossils in a sediment layer
reflect the temperature at the time the sediment layer was formed.
For more details regarding the training sets and reconstruction models,
see \citet{SepBir01}, \citet{Sepetal02} and \citet{refWecetal05}.

The sediment records are supported by chronologies based on multiple
AMS 14C determinations [\citet{SepBir01}; \citet{Sepetal02}]. As the chronology
inevitably contains errors, an attempt is made to take this uncertainty
into account by using the model described in Section~\ref{randomdates}.
Table S.2 in \citet{Eraetal11supp} gives all the data used in our
consensus analysis: the sediment depths, calibrated ages and their
standard errors as provided by the dating laboratory, as well as
pollen-, chironomid- and diatom-based July mean temperature
reconstructions for the lakes Toskal and Tsuolbmajavri.

\subsection{Chronology errors, prebinning}\label{sec3.2}
\label{chronerrors}

\begin{figure}

\includegraphics{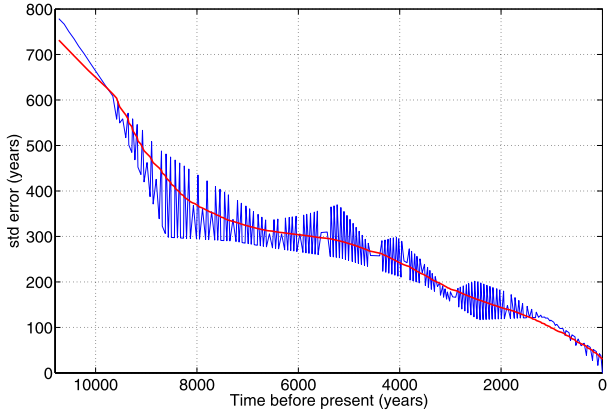}

\caption{Standard errors of the combined binned chronology of the two
sediment cores (blue). Average standard error is plotted when two or
more dates coincide after binning. Also shown is a local linear smooth
that was used in defining the parameters $\psi_i$ of the dating model
likelihood (\protect\ref{datelikelihood}).}
\label{aikahajonnatjasilote}
\end{figure}

The combined chronology (\ref{obsdates}) includes several pairs of
dates with only a few years apart. The spline interpolant used in
representing the consensus temperature anomaly as a continuous function
$\mu(t)$ can exhibit unnatural wiggles between such nearby dates and
we therefore aggregated the dates into 15 year wide bins.
The chronology standard errors of aggregated\vadjust{\goodbreak} dates could then be
averaged, but we actually decided to smooth all of them as shown in
Figure~\ref{aikahajonnatjasilote} and computed the parameters $\psi
_i$ in (\ref{datelikelihood}) from the values of this smooth. It
retains the most important feature of the dating errors, namely, that
they increase considerably when older sediment layers are considered.
These approximations seem reasonable given the large standard errors
associated with the dates and the rather simplistic dating error model
(\ref{datingerrormodel}) used.

\subsection{Priors for reconstruction errors and roughness}\label{sec3.3}
\label{errorsandroughness}
\subsubsection{Reconstruction error}\label{sec3.3.1}
\label{errors}

The prior distribution (\ref{wishartprior}) of $\vecSigma_k$ has the mean
$\EE(\vecSigma_k)=(\nu_k - j_k -1)^{-1}\vecW_k$,
where $j_k$ is the dimension of the $k$th reconstruction $\vecy_k$.
We use a diagonal scale matrix
$\vecW_k = w_k\vecI_{j_k}$ such that $\EE(\vecSigma_k) = \bar
{\sigma}_k^2\vecI_{j_k}$, where $\bar{\sigma}_k^2$
is an estimate for the upper bound of reconstruction error variance.
    Appendix~\ref{errorestimation} suggests  a method to
derive such upper bound estimates and the values obtained are given in
Table~\ref{stdestimatestable}. Since now $\bar{\sigma}_k^2\vecI
_{j_k}=(\nu_k - j_k -1)^{-1}w_k\vecI_{j_k}$,
we must have $w_k/\bar{\sigma}_k^2 = \nu_k -j_k +1$. We set $w_k =
0.5$ for all $k$ which corresponds to degrees of freedom $\nu_k$
between 77.9 and 163.1 and makes the priors rather vague.

The posterior values of the diagonal elements of the matrices
$\vecSigma_k$ turned out to be significantly smaller than their prior
values. As this may suggest that the values
$\bar{\sigma}_k$ are too large (and thus truly only upper bounds),
we also included in our analyses a second set of error covariance
priors by using the value $\bar{\sigma}_k = 0.2$ for all
reconstructions. In this case we opted for a tighter prior by taking
$w_k = 50$ which corresponds to between 1319 and 1410 degrees of
freedom in the priors.

Assuming smaller errors naturally leads to more features in the
consensus analysis being flagged as credible. However, the independent
evidence for some of these features discussed in Section~\ref
{interpretation} can be interpreted as lending some credence to these
smaller reconstruction errors. Trying out different error sizes makes
sense also because it probably is not possible to estimate them very
reliably in the first place. Exploring temperature features for
different error levels could also be thought as a form of scale space
analysis where increasing error levels corresponds to more smoothing.
In the following we refer to these two prior settings as ``large'' and
``small'' errors.

\subsubsection{Roughness}\label{sec3.3.2}
\label{roughness}


The parameter $\lambda_0$ in (\ref{muprior}) is used to describe our
prior belief about the variability or ``roughness'' of the time series
of past temperatures.
In choosing a prior
for $\lambda_0$,
very long instrumental records going back hundreds of years might be useful.
However, the longest records in Finland span only about 150 years, a
period that includes only 2--4 chronology dates for the six
reconstructions considered, thus making roughness estimation impossible.
We therefore decided to use a numerical climate model simulation in
setting the prior roughness level.

\begin{figure}[b]

\includegraphics{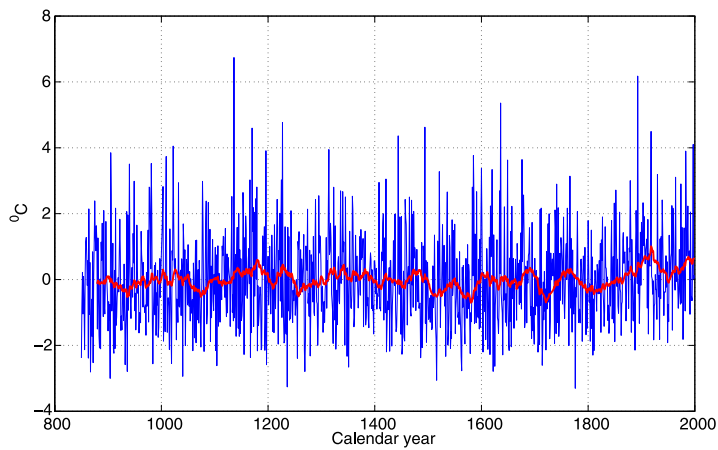}

\caption{Simulated mean July temperature anomaly for Northern Finland
between AD 850 and 1999 (blue curve) together with the 30-year running
mean (red curve).
The vertical axis is the temperature anomaly in centigrade ($^{\circ}$\textup{C}) and
the horizontal axis is the calendar year.}
\label{Ammanplusmean}
\end{figure}

\begin{figure}

\includegraphics{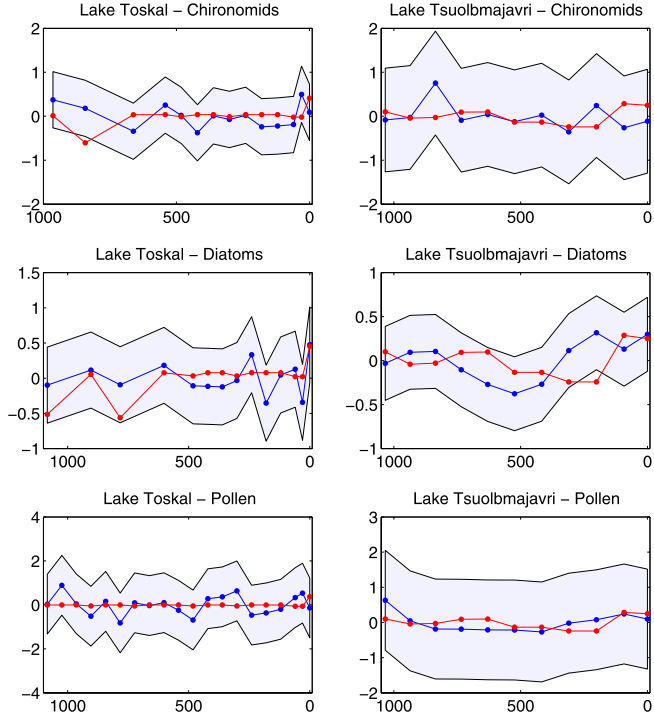}

\caption{The six Holocene mean July temperature reconstructions for
Northern Fennoscandia restricted to the time interval from AD 850 to
1999 (blue curves) together with the simulated 30-year means computed
at the same time points (red curves).
The light blue band around each reconstruction is based on error bars
of size
$\pm2\bar{\sigma}_k$, where the $\bar{\sigma}_k$'s are given
in Table~\protect\ref{stdestimatestable} of the \protect\hyperref[appm]{Appendix}.
The vertical axes show temperature anomaly in centigrade ($^{\circ}$\textup{C}) and
the horizontal axes are time before present in years. Note the
different temperature scales in the figures.}
\label{roughnesscomp}
\end{figure}

A 1150 year long annual mean July temperature series for Northern
Finland, extending from AD 850 to 1999, was extracted from the NCAR
Climate System Model simulation described in \citet{Ammetal07}. The
time series is shown in Figure~\ref{Ammanplusmean} (blue curve). The
six reconstructions should actually be thought of as 30-year averages
of mean July temperatures, sampled at dates included in their
associated chronologies. For visual comparison between the simulation
and the reconstructions we therefore applied a 30-year moving average
to the simulated anomaly (red curve in Figure~\ref{Ammanplusmean}) and
then sampled the average at the dates in the reconstruction chronologies.
The results are shown in Figure~\ref{roughnesscomp}. As one can see,
the reconstructions are at least as rough as the simulation. It
therefore appears that at least some prior smoothing indeed is required
in the consensus analysis which motivates the use of a smoothing prior
(\ref{muprior}) for the consensus. Further, if the simulation is taken
to represent the actual temperature variation, the reconstruction
errors are not very large.
The light blue band around each reconstruction is based on error bars
of size
$\pm2\bar{\sigma}_k$, where the $\bar{\sigma}_k$'s are given
in Table~\ref{stdestimatestable} of the \hyperref[appm]{Appendix}.

\begin{figure}

\includegraphics{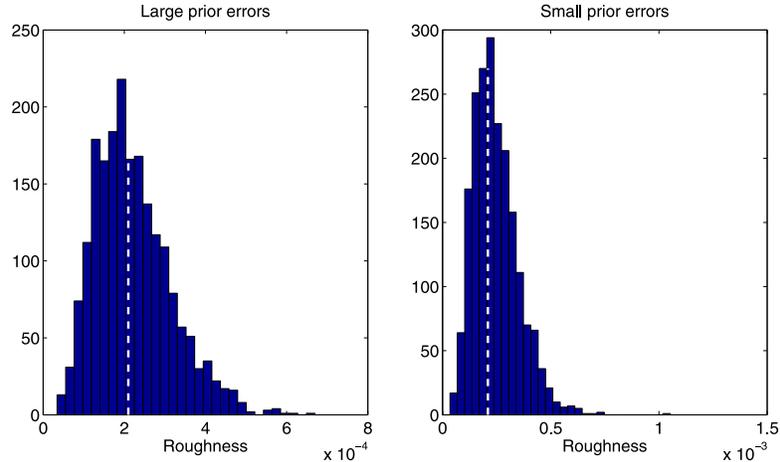}

\caption{Posterior distribution of the roughness measure $R(\vecmu
)=\vecmu^T\vecK\vecmu$ for large (left panel) and small (right)
prior errors. The histograms are based on 2000 sample values and the
dashed line indicates the roughness of the simulation.}
\label{roughnesshist}
\end{figure}

To design a prior for $\lambda_0$, one can use the simulated time
series also for more formal roughness estimation. Given a time series
$\vecmu$, one can
measure its roughness by the quantity $R(\vecmu)=\vecmu^T\vecK\vecmu$
in the exponent of
(\ref{muprior}). For the simulated 30-year running mean, evaluated at
the joint chronology dates (\ref{obsdates}) contained in the interval
from AD 850 to 1999, we have $R(\vecmu) =
2.1 \cdot10^{-4}$. Using the prior $\operatorname{Gamma}(20,0.5)$ for $\lambda
_0$, the posterior mean
of $R(\vecmu)$ is $2.2 \cdot10^{-4}$ and $2.5 \cdot10^{-4}$ for the
large and small prior errors, respectively. In both cases the mean
posterior roughness of the consensus is therefore slightly larger than
that of the simulations which, as indicated in Section~\ref
{fixeddates}, is desirable in order not to smooth too much before scale
space analysis is carried out. We therefore used $\operatorname{Gamma}(20,0.5)$
as the prior distribution for $\lambda_0$. Figure~\ref{roughnesshist}
shows the posterior distribution of $R(\vecmu)$ for both large and
small prior error settings with the roughness of the simulation
depicted as a dashed line. By testing other reasonable alternatives we
also concluded that neither the mean nor the width of the prior
distribution of $\lambda_0$ has a major effect on the estimated
consensus features.

\begin{figure}

\includegraphics{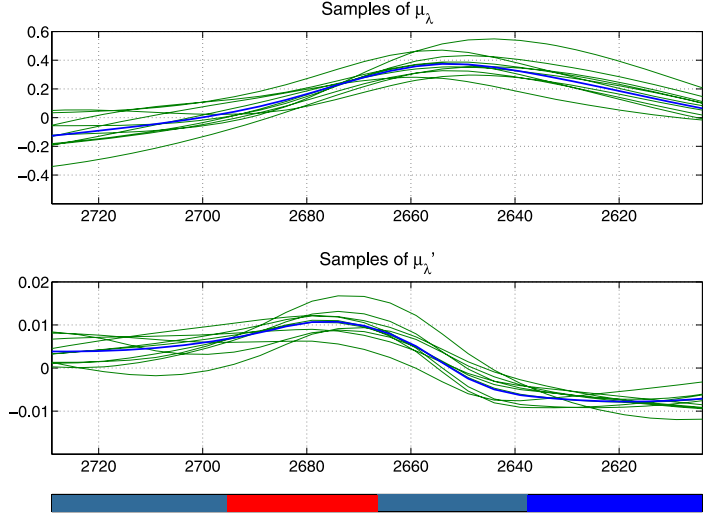}

\caption{Upper panel: sample curves of $\mu_{\lambda}$ (green)
together with the posterior mean $\EE(\mu_{\lambda} | \vecy,\vect
)$ (blue). Lower panel: corresponding samples of
$\mu_{\lambda}'$ and the posterior mean $\EE(\mu_{\lambda}' |
\vecy,\vect)$. The color bar on the bottom depicts posterior sample
based inference on the sign of $\mu_{\lambda}'$. For more
information, see the text.}
\label{realizations}\vspace*{-6pt}
\end{figure}

\begin{figure}[t!]

\includegraphics{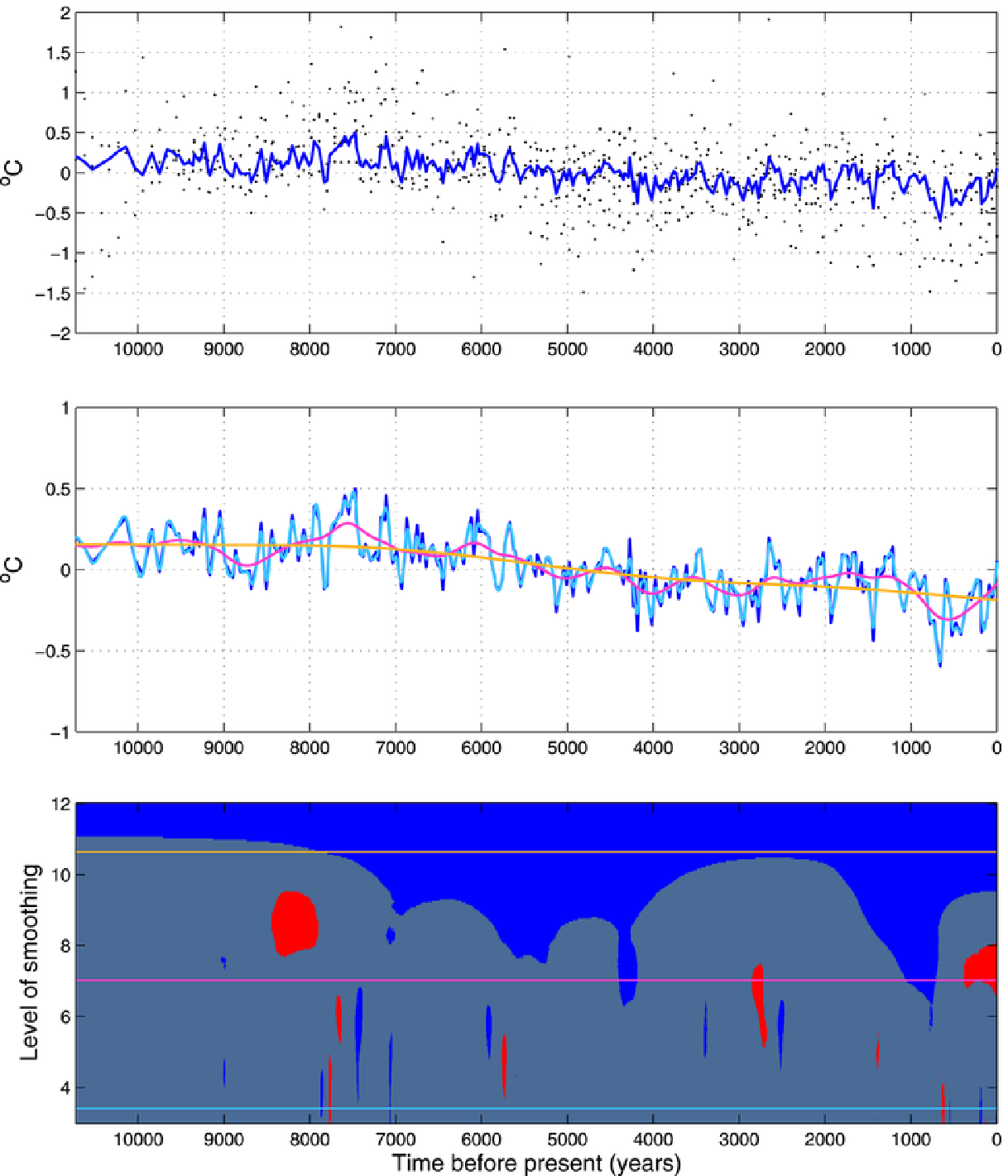}

\caption{Scale space analysis of the consensus of six temperature
reconstructions.
The top panel shows the reconstructions (dots) and the posterior mean
of the consensus (blue curve). Large reconstruction errors were assumed
and the credibility level
$\alpha= 0.8$. The middle panel shows the posterior mean of the
consensus together with three smooths of the posterior consensus
corresponding roughly to multi-decadal (light blue), centennial
(purple) and millennial (yellow) time scales. The bottom panel is the
credibility map where blue and red indicate credible cooling and
warming, respectively. For more information see the text.}
\label{3panelanalysisTR}\vspace*{-6pt}
\end{figure}

\subsection{The consensus and its credible features}\label{sec3.4}

\begin{figure}

\includegraphics{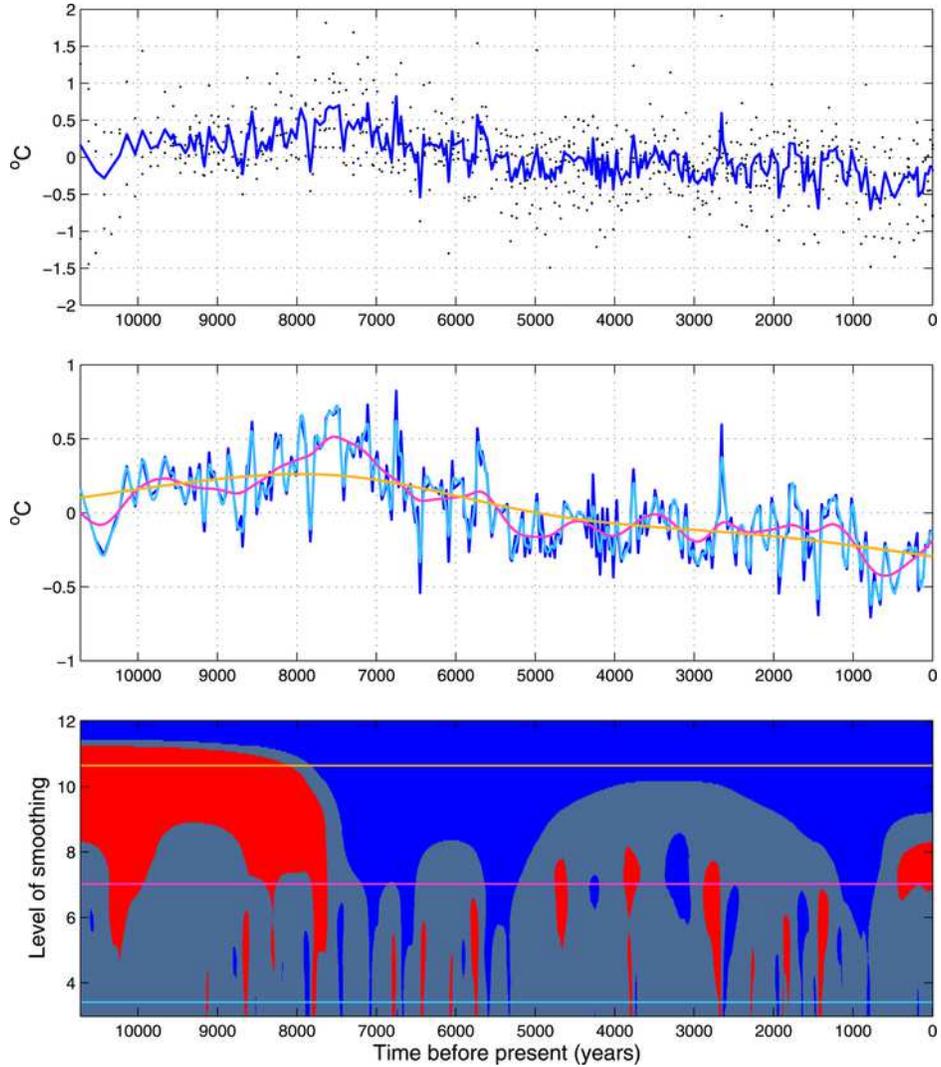}

\caption{Scale space analysis of the consensus of six temperature
reconstructions.
Small reconstruction errors were assumed and the credibility level
$\alpha= 0.8$. For more information see the caption of Figure~\protect\ref
{3panelanalysisTR} and the text.}
\label{3panelanalysis004}\vspace*{6pt}
\end{figure}

%
Scale space analyses of the consensus anomaly with large and small
prior reconstruction errors are shown in
Figures~\ref{3panelanalysisTR} and~\ref{3panelanalysis004}, respectively.
The top panel shows the reconstructed temperature anomalies (dots)
together with the posterior mean of the consensus (blue curve).
The middle panel shows the posterior mean again together with three
smooths $\EE(\mu_{\lambda} | \vecy,\vect)$ of the posterior
consensus corresponding roughly to multi-decadal (light blue),
centennial (purple) and millennial (yellow) time scales (cf.
Section~\ref{featureanalysis}).
Comparing with the ad hoc methods discussed in the \hyperref[sec1]{Introduction}, we
observe that there is a qualitative correspondence between the
smoothing based curves of Figure~\ref{simplemethods} (red and green
curves) and the centennial level posterior means of our scale space
analyses as well as between the mean of the spline interpolants (lower
panel, blue curve) and our multi-decadal posterior mean.

The bottom panel is a feature credibility map where the vertical axis
represents the smoothing level $\lambda$ (in logarithmic units), that
is, the time scale at which the features are examined. The smoothing
levels corresponding to
the three smooths of the middle panel are indicated by horizontal lines
of the same color. A pixel at a location $(s_j,\lambda)$ is colored
blue or red depending on whether the slope of the smoothed anomaly $\mu
_{\lambda}$ is credibly negative or positive. Thus, blue and red
indicate cooling and warming, respectively, at the particular time
$s_j$ and scale $\lambda$ considered. Flagging of negative and
positive slopes is based on their joint posterior probability which is
required to exceed a given threshold $\alpha$, typical values used
being in the range [0.8, 0.95]. Gray color indicates that the sign of
the slope is not credibly different from zero.

Figure~\ref{realizations} is a schematic illustration of how the map
is drawn, focusing on the interval from 2729 to 2604 years before
present and a multi-decadal smoothing level~$\lambda$. In the upper panel, a few sample curves of $\mu_{\lambda
}$ (green) together with the posterior mean $\EE(\mu_{\lambda} |
\vecy,\vect)$ (blue) are shown. The lower panel shows the
corresponding samples of
$\mu_{\lambda}'$ and the posterior mean $\EE(\mu_{\lambda}' |
\vecy,\vect)$. The color bar on the bottom depicts posterior sample
based inference for the chosen fixed value of $\lambda$, where, with
posterior probability at least $\alpha$, the derivative of $\mu
_{\lambda}$ is positive or negative on the intervals indicated by red
and blue, respectively, and the probability is computed\vadjust{\goodbreak} jointly over
all time points $s_j$ in these intervals. The full map, such as in the
middle panels of Figures~\ref{3panelanalysisTR} and \ref
{3panelanalysis004}, is obtained by stacking such color bars, for the
whole Holocene and for all scales $\lambda$ considered.

As in our earlier scale space analyses of the paleoclimate, the
credibility level was
chosen as $\alpha=0.8$ [e.g., \citeauthor{refEraHol03} (\citeyear{refEraHol03,refEraHol05,refEraHol05b}); \citet{refWecetal05}].
Increasing the level, say, to 0.95, slightly shrinks the credible
features (blue and red areas)
but does not affect much the interpretation given in Section~\ref
{interpretation}. The
$\alpha=0.95$ versions of all consensus credibility maps are included
in the supplement
[\citet{Eraetal11supp}].

\begin{figure}

\includegraphics{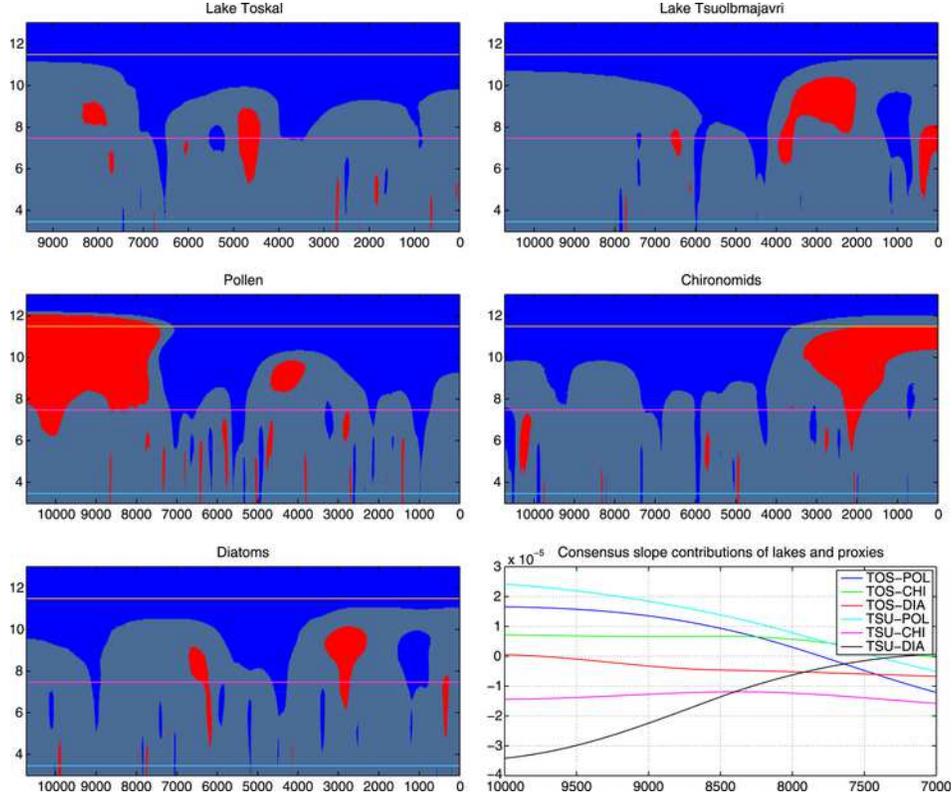}

\caption{Consensus based on subgroups of the six temperature
reconstructions considered. Large reconstruction errors are assumed and
the credibility level is 0.8.
In the top row, the Lake Toskal map is based on all three proxy records
obtained from that lake and similarly for Lake Tsuolbmajavri.
The other three
maps show the consensus according to each proxy when the corresponding
proxy records from each lake have been combined.
The bottom panel of the second column is a more detailed analysis of
how each reconstruction affects the overall consensus within a
particular time interval on a millennial time scale.
For more information see the caption of Figure~\protect\ref{3panelanalysisTR}
and the text.}
\label{anatomy}
\end{figure}

It is interesting to study also the effects on the consensus of the two
lakes and the three proxies separately. Such an analysis is presented
in Figure~\ref{anatomy}, where credibility maps for the lakes and the
proxies based on large reconstruction errors are
displayed.
One can also analyze the role of each of the six reconstructions more
quantitatively by considering their mean contributions to the posterior
consensus.   Appendix~\ref{weightanalysis} proposes such
an approach and to demonstrate the idea, we examined more closely the
early Holocene warming suggested in the credibility map of Figure~\ref
{3panelanalysis004}.
The bottom panel of the second column of Figure~\ref{anatomy} shows
the mean contribution of each reconstruction to the slope of the
consensus at a millennial time scale (yellow curve in Figure~\ref
{3panelanalysis004}), from the beginning of the Holocene to 7000 years
before present. Such a plot can be useful when one wants to focus the
analysis on a particular feature in a limited time window.

The results of Figures~\ref{3panelanalysisTR}--\ref{anatomy} are
based on $\vecmu$-samples of size 4000 where the first 2000 were used
for burn-in. Generating such a sample on a standard PC takes about 10
hours. A uniform grid of about 2000 time points $s_j$ and a logarithmic
grid of 200 smoothing levels $\lambda$ were used in the scale space
analyses. With random dates it takes about 10 hours to process a batch
of 10 smoothing levels. Computations can be sped up by allocating the
batches to different processors. Parameter convergence was checked
visually. Initial values were picked from the priors for those
parameters that are updated by Gibbs sampling and the carbon dating
based values were used to initialize the chronologies. The posterior
error covariances were almost diagonal but heteroskedastic with small
off-diagonal elements.
The chronologies changed only little in the simulation. The standard
error of a radiocarbon date is commonly interpreted as a standard
deviation of a normal distribution center at the date [cf. (\ref
{datelikelihood})].
To test the robustness of dating error assumptions, we repeated some of
our analyses assuming either a much smaller (down to zero) or a much
larger (up to several times the value used in the reported analyses)
standard error, but the features proposed by the maps stayed the same.
For very large standard errors this is due to proposals in the MCMC
simulation being mostly rejected.

\subsection{Interpretation of results}\label{sec3.5}
\label{interpretation}

\subsubsection{Consensus features}\label{sec3.5.1}
\label{consensusfeatures}

According to the credibility maps of Figures~\ref{3panelanalysisTR}
and~\ref{3panelanalysis004}, overall cooling is the longest time scale
feature of Holocene summer temperature in northern Finland, indicated
by the continuous blue color in the topmost part of the maps.
This is thought to be mostly due to the earth's changing orbital
geometry around the sun.
At millennial scales (yellow lines in the maps), the consensus summer
temperatures exhibit some other key aspects of Holocene climate
evolution, such as an early Holocene warming trend shown strongly in
Figure~\ref{3panelanalysis004} and weakly in Figure~\ref
{3panelanalysisTR}, together with a peak warming at around 8 kyr BP
(8000 years before present) indicated by red changing to blue, followed
by a monotonic cooling trend (blue color) until the present time. This
overall pattern is predominantly driven by annual mean and summer
orbital forcing at the high northern latitudes [\citet{BerLou91}]. In
the Northern Hemisphere summer months the incoming solar radiation
(insolation) peaked between 11 and 9 kyr BP [\citet{Kut81}], when
insolation was approximately 7--9\% higher than at present at
$70^\circ\,\mathrm{N}$, and gradually declined since then. The relatively
cool summer temperatures in the early Holocene (rising trend before 9
kyr BP) in the consensus hence refer to a slightly delayed timing of
the Holocene Thermal Maximum (HTM) relative to this peak summer
insolation, suggesting that the climate response to the orbital forcing
must also be affected by some extra forcings and internal feedbacks in
the climate system [\citet{Chaetal2000}]. The cool conditions in the
earliest Holocene were apparently heavily influenced by the last
substantial remnants of the large Fennoscandian and Laurentide
continental ice sheets that trigged changes in ocean heat
transportation and surface albedo [\citet{KapWol06}; \citet{Renetal09}].

According to our consensus reconstruction, HTM in northern continental
Europe occurred at around 8--9 kyr BP, when the inferred summer
temperature values clearly exceeded the modern levels. This early
peaking of Holocene warmth
contradicts several earlier studies that place the timing of peak
warming across a wide area of northern Europe closer to mid-Holocene at
around 6 kyr BP
[\citet{Davetal03}; \citet{Macetal2000}; \citet{Kaufetal04}]. Evidence for the
mid-Holocene thermal maximum in northern Europe comes largely from a
northward and upward expansion of northern treelines, as well as from
retreating glaciers [\citet{Janetal2007}]. However, a recent global
assessment of treeline response to climate warming suggests that
treeline advance may be more strongly associated with winter, rather
than summer, warming [\citet{Haretal09}]. In addition, in many parts of
Scandinavia, glaciers started to retreat in the early Holocene, soon
after the transient cooling event, termed the Finse event [8.5--8.0 kyr
BP; \citet{Nesetal08}]. The early expression of peak summer warming
identified in the present study is further consistent with a recent
model simulation study [\citet{Renetal09}], where maximum summer warmth
in the northeast of Europe was placed closer to 8 kyr BP.


At multi-decadal to centennial scales (light blue and purple lines in
the maps), climate variability as highlighted in our small-error
analysis (less so with large reconstruction errors) shows a complex
picture with indications of repeated warm and cold climate episodes,
the specific causes of which are not fully understood. Some of the
peaks found in our record seem to be coherent with the Holocene series
of North Atlantic ice-rafting events defined by \citet{Bonetal97}
within the dating uncertainties ($\pm$100 to 200 years). These include
the weak temperature minima
at around 1.4, 2.8, 4.2 and around 10.3 kyr BP, whereas the remaining
mid- and early Holocene ``Bond events'' are not evident in our record.
Neither can we find any event-like feature around the classical 8.2 kyr
BP cooling event [\citet{Alletal97}], although the most pronounced
decline in overall Holocene summer temperatures started in our record
around this time (see above). Examination of the maps at the smallest
smoothing levels shows credible fluctuations in summer temperature, in
particular, between 7.0 and 5.0 kyr BP and from 3.0 kyr BP to the
present, while more stable conditions occurred between 5.0 to 3.0 kyr
BP and in the early Holocene. Solar variability is the most plausible
explanation for the temporal dynamics of these short-term changes.
Indeed, recent work utilizing spectral analysis of radionuclide records
suggests that the solar cycles were particularly prominent during the
time intervals 6.0--4.5 kyr BP and 3.0--2.0 BP, whereas this periodic
behavior faded during other time intervals [\citet{Knuetal09}]. Hence,
the high-variability intervals in our record coincide with the periods
of intensive solar cycles, which in turn correlate with periods of
significant reorganization of the ocean and atmospheric circulation in
the North Atlantic region
[\citet{Mayetal04}; \citet{Seietal07}].

Our scale space consensus analysis (in particular, the credibility map of
Figure~\ref{3panelanalysis004}) indicates that the Northern
Fennoscandia summer climate experienced a succession of warming and
cooling events during the most recent part of the Holocene, broadly
similar to those documented earlier in Northern Hemisphere temperature
reconstructions, including the Current Warm Period (CWP), Little Ice
Age (LIA) and Medieval Climate Anomaly (MCA) [\citet{Janetal2007}; \citet{Manetal08}]. The MCA commenced around 1.3 kyr BP and
terminated around 0.8 kyr BP when temperatures started to decrease
toward the LIA. Conditions slightly warmer than those of the 20th
century may have prevailed in the North Atlantic climate regime during
the MCA as deduced on the basis of our analysis. The peak medieval
warmth is around 1.2 kyr BP in our record, which is earlier than in
many previous published reconstructions, but is in accordance with
\citet{Manetal08} who place the MCA between AD 1450 and AD 700. The LIA
in our consensus reconstruction occurred perhaps between ca. 0.5 and
0.15 kyr BP (about AD 1500--1850), in agreement with the recent
Arctic-wide synthesis of proxy temperature records [\citet{Kaufetal09}].
%
%
The recent warming (CWP) shows as a credibly positive temperature trend
in centennial scales.

\subsubsection{Contributions from the proxies and the lakes}\label{sec3.5.2}
\label{anatomyanalysis}

Looking at the lake- and proxy-specific credibility maps of Figure~\ref
{anatomy}, we note first that, of the three proxies, the pollen-based
reconstructions suggest most features with somewhat fewer credible
features exhibited by the chironomid and the diatom records. All three
agree on a Holocene-wide cooling trend which therefore becomes part of
the overall consensus. Still, on millennial scales (yellow line), the
cooling trend after about 4 kyr BP in the chironomid record is a bit
less certain than in the two other proxies. It is notable that evidence
for early Holocene warming
and the HTM in the overall consensus appears to come from the pollen
record only.
The millennial scale detail analysis shown in the bottom panel of the
second column of
Figure~\ref{anatomy} clearly confirms this.
The fact that in the large-error analysis of
Figure~\ref{3panelanalysisTR} these show only weakly is probably due
to the relatively large pollen reconstruction error upper bounds used
for this analysis
(cf. Table~\ref{stdestimatestable}). The LIA is clearly visible as a
credible temperature minimum only in the diatom record. However,
combined with the cooling trend immediately prior to it, which is
present also in pollen and chironomid reconstructions, the LIA signal
in diatoms is strong enough to show in the consensus, too. The Bond
events (cf. Section~\ref{consensusfeatures}) are supported in varying
degrees by different proxies. The warm
MCA appears to be better reconstructed by
chironomids than pollen.
The recent centennial-scale rise in temperatures exhibited in the
consensus is driven mostly by the diatom record with the chironomids
showing millennial scale warming during the last 2000--3000 years.

Considering the credibility maps in the first row of Figure~\ref
{anatomy}, we notice that
the records from the two lakes both support overall Holocene cooling
and the LIA (although only barely for Toskal), whereas only Lake Toskal
shows weak evidence for early Holocene warming. In light of the detail
analysis of Figure~\ref{anatomy} (lower right-hand corner panel), it
appears that the strong millennial scale warming signal in the Lake
Tsuolbmajavri pollen record is drowned by negative contributions from
the chironomid and diatom reconstructions.
Still, as noted above, when evidence in all records is included, the
warming signal is strong enough to show in the overall consensus.
Finally, we observe that only the Lake Tsuolmbajavri record suggests
the presence of the MCA and that
opposite features in the lake records at around 4 kyr BP may be the
source of centennial-scale oscillations in the consensus during 5--3
kyr BP (purple curve in the middle panel of Figure~\ref{3panelanalysisTR}).

\section{Discussion}\label{sec4}
\label{discussion}

Given a collection of noisy reconstructions, the proposed method
uses Bayesian inference to find those features of past climate
variation that are supported by their consensus. Although only
temperature was considered, other climate variables could be handled similarly.
Further, while the reconstructions considered in this paper were based
on radiocarbon dated sediments samples, the method is conceivably
applicable to other proxy types that use different dating methods such
as tree rings, varved lake sediments, ice cores and speleothem
archives, where estimates of dating errors are available [see \citet{Jonetal09} for a discussion of these and other proxy types]. In case
of annually resolved records such as tree rings, the fixed dates
version of the method might suffice. Also, although the paper focuses
on an application to paleoclimate reconstruction, the
method developed is likely to find use also in other contexts where a
combination of
information across several noisy time series is of interest.

Handling of dating errors in our consensus model could probably be
considerably improved.
A sophisticated Bayesian dating error model, BChron, was introduced in
\citet{HasPar08}. Other recent proposals include, for example,
\citet{Bla03}, \citet{BlaChr05} and \citet{BroRam07}. The problem of
modeling the relationship between sediment depth and age was also
analyzed in \citet{Teletal04} and \citet{Heeetal05}, and aligning
multiple varve chronologies\vadjust{\goodbreak} was considered in \citet{Aueetal08}.
Dating error models developed for spatial problems could also be useful;
see, for example, \citet{FanDig11} and \citet{CreKor03}.
Still, while we readily acknowledge that the error model described in
Section~\ref{randomdates} may be too crude to reflect all aspects of
uncertainty in the dating process, it nevertheless can serve as a first
approximation that allows, in principle, the effect of dating errors to
enter the posterior uncertainty of the consensus anomaly. In future
work we hope to incorporate in the analysis ideas from more
sophisticated error models such as Bchron. Such an improvement in the
analysis might be incorporated also in a system that uses Bayesian
reconstructions to begin with.
We leave these ideas for future work.

Another direction of development would be to include the spatial
dependencies between the proxy records in the model. With only two core
locations considered in our example, this is not relevant, but it might
be useful when more locations are included in the consensus analysis.

We proposed to use climate simulations to gain insight into the
variability of the past temperature.
Of course, the simulation we used covers only a fraction of the
approximately 10\mbox{,}000 years considered in the reconstructions and,
therefore, in the analyses described in Section~\ref{roughness}, one
considers temperature roughness only for about 10\% of the whole
Holocene period. Still, although the mean temperature levels for the
last 1150 years may be different from those during the rest of the
Holocene, it may not be unreasonable to assume that the inter-annual
temperature variation has not changed dramatically. By studying the
simulated 30-year mean for the last 1150 years we may therefore gain at
least some idea of its roughness during the whole Holocene. In a sense,
such an assumption could be viewed as being somewhat analogous to the
basic premise underlying proxy-based paleoclimate reconstructions,
namely, that the relationship between the proxy records and the climate
has not changed over thousands of years.


To summarize, the method described in this paper provides a means to
estimate the consensus temperature variation in heterogenic time series
and also to capture its salient features, such as maxima, minima and
trends in different time scales in a statistically principled manner.
Our model allows dating uncertainties, distinct or overlapping core
chronologies, as well as time-varying, correlated reconstruction errors
that can also have different magnitudes for different proxies and
cores. We believe that the method has also wider applicability
potential in data mining of various types of climate records and
compiled time series. When applied to lake data series from northern
Finland, a millennial-scale cooling trend was found since the Holocene
thermal maximum at around 8 kyr BP associated with the decrease in
orbitally driven summer insolation. Superimposed on the
millennial-scale trends, the summer climate in northern Finland was
punctuated by several quasicyclical climate events, the forcing
mechanisms of which are not yet fully understood. Our scale space
analysis also suggests that inconsistencies in climate reconstructions
and their interpretations may be at least partly spurious; there is
probably no single narrative that counts as the canonical version of
Holocene climate change. Instead, there are many interpretations
depending on the proxy and the resolution at which the data are gained
and examined.
Finally, while the paper focuses on paleoclimate time series, the
proposed method can be applied in other contexts where one seeks to
infer features that are jointly supported by an ensemble of irregularly
sampled noisy time series.

\begin{appendix}\label{appm}
\section{Estimation of the reconstruction error}\label{secB}
\label{errorestimation}


We explain here how the temperature anomalies $\vecy_k$ were used to
estimate upper bounds for the reconstruction error variances.

\begin{table}[b]
\tabcolsep=0pt
\tablewidth=180pt
\caption{Estimates of upper bounds of reconstruction errors for the 6
proxy records considered}
\label{stdestimatestable}
\begin{tabular*}{180pt}{@{\extracolsep{\fill}}lc@{}}
\hline
\textbf{Proxy record} & $\boldsymbol{\bar{\sigma}_k}$\\
\hline
Lake Toskal chironomids & 0.32\\
Lake Toskal diatoms & 0.27 \\
Lake Toskal pollen & 0.68 \\
Lake Tsuolbmajavri chironomids & 0.59 \\
Lake Tsuolbmajavri diatoms & 0.21 \\
Lake Tsuolbmajavri pollen & 0.71 \\
\hline
\end{tabular*}
\end{table}

Assuming that $\vecy_k \sim\N(\vecmu_k,\sigma_k^2\vecI_{j_k})$,
the distribution of the random variable $V_k = \|\vecy_k\|^2 = \vecy
_k^T\vecy_k$ is determined
by the parameter $\vectheta_k = (\vecmu_k,\sigma_k)$.
We consider a fixed value $\bar{\sigma}_k > 0$ and the null hypothesis
\[
H_0\dvtx  \Theta_0 = \{\vectheta_k = (\vecmu_k,\sigma_k) \mid
\vecmu_k \in\matR{m}, \sigma_k \geq\bar{\sigma}_k\}
\]
against the alternative
\[
H_1\dvtx  \Theta_1 = \{\vectheta_k = (\vecmu_k,\sigma_k) \mid
\vecmu_k \in\matR{m}, \sigma_k <\bar{\sigma}_k \}.
\]
The null hypothesis is rejected if $V_k \leq\bar{v}_k$, where $\bar
{v}_k$ is
some fixed value. It is shown in \citet{techrepHolEra01} that the
significance level of
this test is given by
%
\begin{equation}
\label{siglevel}
\beta= \PP(\chi^2_{j_k-1} \leq\bar{v}_k/\bar{\sigma}_k^2),
\end{equation}
where $\chi^2_{j_k-1}$ is a chi-square variable with $j_k-1$ degrees
of freedom.
Setting $\beta= 0.05$, an upper bound for $\sigma_k$ can therefore be
estimated as
\[
\bar{\sigma}_k = \sqrt{V_k/\chi^2_{j_k-1,0.05}},
\]
where $\chi^2_{j_k-1,0.05}$ is the
5th percentile of the $\chi^2$-distribution with $j_k-1$ degrees of freedom.
These values are listed in Table~\ref{stdestimatestable} for the six\vadjust{\goodbreak}
proxy records and they were used to define the large-error prior scale
matrices $\vecW_k$ in the consensus analysis.

\section{Contributions of individual proxy records to the consensus}\label{secC}
\label{weightanalysis}

It follows from (\ref{likelihood1}) and (\ref{muprior}) that
\[
\vecmu| \{\vecSigma_k\},\lambda_0, \vectau, \vecy, \vect  \sim
  \N(\vecmu_0,\vecSigma_0),
\]
where
\[
\vecSigma_0 =  \Biggl(\sum_{k=1}^m \vecSigma_k^{-1} + \lambda_0\vecK
 \Biggr)^{-1}
\]
and
\[
\vecmu_0 = \vecSigma_0 \Biggl(\sum_{k=1}^m \vecSigma_k^{-1}\vecy
_k \Biggr) =
\sum_{k=1}^m \vecSigma_0\vecSigma_k^{-1}\vecy_k,
\]
where it is understood that $\vecSigma_k$ and $\vecy_k$ are extended
to an $n \times n$
matrix and an $n$-dimensional vector, respectively, by putting zero
entries to locations that correspond to those time points in the full
joint chronology
$\vect$ that do not appear in the chronology $\vect_k$ of proxy
record\vspace*{1pt} $k$. It follows that the components of the posterior mean vector
$\EE(\vecSigma_0\vecSigma_k^{-1}\vecy_k | \vecy,\vect)$ can be
used to quantify the contribution of record $k$ to the posterior of
$\vecmu$ at the time points $\tau_1,\ldots,\tau_n$. If $\vecS
_{\lambda}$ and $\vecD$ are the matrices
defined in Section~\ref{featureanalysis}, the contribution of record
$k$ to the slope of the smooth
$\mu_{\lambda}'$ at the time points $s_1,\ldots,s_r$ [cf. (\ref
{derivatives})] can then be analyzed by considering the mean of
$\EE(\vecD\vecS_{\lambda}\vecSigma_0\vecSigma_k^{-1}\vecy_k |
\vecy,\vect)$, instead. This is the quantity depicted for each
reconstruction in the bottom panel of the second column of Figure~\ref
{anatomy}.
\end{appendix}

\section*{Acknowledgment}
We are grateful to Dr. Caspar Ammann from NCAR who provided us with the
simulated temperature times series used in Section~\ref{roughness}.

\begin{supplement}[id=suppA]
\sname{Supplement A}
\stitle{Additional analyses and the data used\\}
\slink[doi]{10.1214/12-AOAS540SUPPA} 
\slink[url]{http://lib.stat.cmu.edu/aoas/540/Supplement_A.pdf}
\sdatatype{.pdf}
\sdescription{The document (a pdf-file) reports exploratory analyses
of the estimated reconstruction errors,
shows additional credibility maps, and provides the data analyzed in
the article.}
\end{supplement}

\begin{supplement}[id=suppB]
\sname{Supplement B}
\stitle{The Matlab code}
\slink[doi]{10.1214/12-AOAS540SUPPB} 
\slink[url]{http://lib.stat.cmu.edu/aoas/540/Supplement_B.zip}
\sdatatype{.zip}
\sdescription{The Matlab code (in a zip-file) used to compute the
results of the article.}
\end{supplement}


\printaddresses

\end{document}